\documentclass[a4paper,pra,preprint,floatfix,superscriptaddress]{revtex4}
\usepackage[english]{babel}
\usepackage{amsmath,amsfonts,amssymb,float}
\usepackage{graphicx}
\usepackage{dcolumn,bm}
\usepackage{verbatim} % ADDED TO USE COMMENT BLOCKS MAY NEED REMOVED
\usepackage{color}
\usepackage{url}

\DeclareMathOperator{\sinc}{sinc}

\def\clf{Central Laser Facility, STFC Rutherford Appleton Laboratory, Didcot, OX 11 0QX, United Kingdom}
\def\strathclyde{Department of Physics, SUPA, University of Strathclyde, Glasgow, G4 0NG, United Kingdom}

\def\imperialchem{Department of Chemistry, Molecular Sciences Research Hub, Imperial College London, SW7 2AZ London, UK}
\def\icmm{Instituto de Ciencia de Materiales de Madrid (ICMM), Consejo Superior de Investigaciones Cient{\'\i}ficas (CSIC), 28049 Madrid, Spain}
\renewcommand{\vec}[1]{\ensuremath{\mathbf{#1}}}

\begin{document}

\title{A beat wave approach to harmonic generation in chiral media}

\author{Raoul Trines}
\email[Corresponding author, ]{raoul.trines@stfc.ac.uk}
\author{Holger Schmitz}
\affiliation\clf
\author{Robert Bingham}
\affiliation\clf
\affiliation\strathclyde
\author{Martin King}
\author{Paul McKenna}
\affiliation\strathclyde
\author{David Ayuso}
\affiliation\imperialchem
\author{Laura Rego}
\affiliation\icmm
\date\today

\begin{abstract}
%In previous studies on laser harmonic generation in plasma and solids, we have developed a beat-wave approach to laser harmonic generation, in which all Fourier spectra take the form of regular grids. In this paper, we extend our model to the nonlinear optical response of isotropic chiral media driven by locally chiral light, in which the tip of the electric-field vector draws a chiral Lissajous figure in time. This includes chiral high harmonic generation driven by synthetic chiral light that is globally chiral, exhibits polarisation of chirality, or is topologically structured in space, as well as enantio-sensitive imaging in the perturbative regime. As in our earlier work on laser-solid interactions, the medium is represented by a zero-frequency (DC) driving mode. We show how an enantio-sensitive DC mode can be derived from the interaction of synthetic chiral light with a chiral medium. The beating between this DC mode and the EM fields then leads to a regular harmonic spectrum with alternating chiral and achiral modes. We will derive the criteria for these modes to overlap in Fourier space, so they can combine to yield enantio-sensitive interference patterns or line intensities. Finally, we will apply our framework to a variety of existing results to validate its predictions.
We extend the beat-wave framework for laser harmonic generation -- where spectra form regular lattices in Fourier space -- to the nonlinear response of isotropic chiral media driven by locally chiral light. We represent the enantio-sensitive response of the medium by a chiral zero-frequency (DC) mode derived from the transverse spin density induced by structured or focused fields. Beating between this DC mode and the driving electromagnetic modes yields alternating chiral and achiral contributions on a regular harmonic lattice. We derive a general criterion for when chiral and achiral pathways overlap at the same harmonic and generate enantio-sensitive interference that survives spatial or angular integration (global chirality), versus when enantio-sensitivity remains confined to spatially varying patterns (local chirality). We apply the criterion to published configurations of synthetic chiral light, including OAM-carrying bicircular fields and crossed multicolour beams, and show that it reproduces and clarifies their reported global-chirality and beam-bending regimes.

\end{abstract}

\maketitle

\section{Introduction}

%Why is harmonic generation in chiral media so important? What has been done before? How does that fit with the current work? Leave this mostly to Laura Rego and David Ayuso, as they are the experts.

% Synthetic chiral light enables highly efficient enantio-sensitive nonlinear optics within the electric-diipole approximation by encoding handedness locally in the sub‑cycle trajectory of the electric‑field vector. 

Chirality is the property of an object that cannot be superimposed to its mirror image. The universality of chirality allows us to find it across a variety of fields, from fundamental physics to medicine, and scales, from subatomic particles to galaxies. It is especially relevant in chemistry and in the pharmaceutical industry \cite{francott, brooks}, as chiral molecules are commonly present in many chemical and biological processes. In the case of chiral molecules, the two versions of a molecule (mirror images of each other) are called enantiomers, usually referred as right-handed and left-handed enantiomers, in analogy with the chirality of our hands. Opposite enantiomers can behave very differently when interacting with another chiral object, such as other chiral molecules. Thus, they can interact very differently with living organisms, which makes their distinction crucial.

Efficiently distinguishing between chiral molecules is, however, challenging. A widely used tool for enantiomer discrimination is circularly polarised light, as its electric field draws a helix in space, which is a chiral object. However, typical linear interactions with circularly polarised light lead to less than 0.1\% enantio-sensitive signals in chiral dichroism experiments, where the difference in absorption of light with opposite handedness is measured \cite{berova}. Similarly, optical activity, another classic measurement of chirality where linearly polarised light acquires opposite tilt angles upon propagation through samples with opposite enantiomers, results in equally weak differential signals. The origin of this lack of sensitivity resides in the fact that, in such measurements, the helix drawn by the light is several orders of magnitude larger than the size of the chiral molecules. In other words, the molecules need to interact with light beyond the dipole approximation, thus responding to the local variations of the electric field (i.e., interacting with the magnetic component of the light).

\begin{figure}[t]
        \includegraphics[width=\columnwidth]{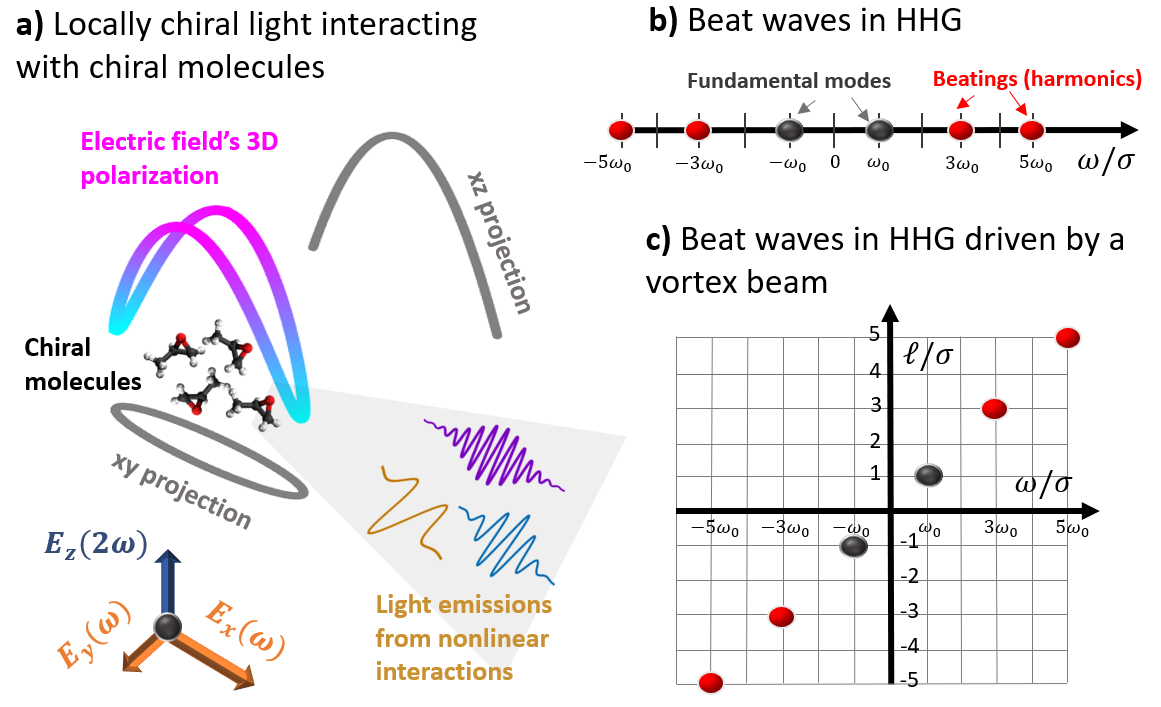}
\caption{(a) Illustration of chiral molecules interacting nonlinearly with an electric field with local 3D polarisation, which is composed of an elliptically polarised $\omega$ field in the $(x,y)$ plane (orange arrows) and a linearly polarised $2\omega$ field in the $z$-direction (blue arrow). (b) Beat-wave approach reproducing the generation of harmonics (red dots) as beatings of the fundamental modes (grey dots) of a linearly polarised driving field. (c) Beat-wave approach applied to harmonic generation driven by a linearly polarised vortex beam with $\ell_0 =1$. The typical OAM scaling law, $\ell_q=q \ell_0$ where $q$ is the harmonic order, is reproduced.}
        \label{figure1}
\end{figure}

One of the solutions to this problem that has been explored during the last decade is the so-called ``electric-dipole revolution'' \cite{ayuso3}: finding methods that rely solely on electric-dipole interactions. This is possible if one creates light that is locally chiral: the polarisation of light's electric field draws a 3D Lissajous figure in time at each point of space. Thus, the chiral molecules interact with a chiral light within the dipole approximation. In order to create such type of light, also known as synthetic chiral light, one needs two ingredients: a longitudinal polarisation component (obtained in non-collinear combinations of beams or tight-focused beams \cite{lax}) and at least two frequencies. Therefore, molecules need to interact with both frequencies to interact with the whole locally chiral field, which brings the necessity of nonlinear interactions, see Figure \ref{figure1}(a).

In recent years, a compilation of methods to investigate chirality based on nonlinear interactions driven by locally chiral light has been developed, including non-collinear configurations resulting in high-order harmonic generation (HHG) \cite{ayuso1, ayuso2, rego1}, sum-frequency generation \cite{vogwell1} or topological chiral light \cite{mayer1}, among others \cite{fischer}. This merging between two prominent fields, chirality and nonlinear optics, raises the following question: which tools from nonlinear optics can we bring to the realm of chirality?

In the study of synthetic chiral light interacting with chiral media, two complementary regimes have been emphasized: (i) global chirality, where enantio-sensitivity survives integration over transverse coordinates and is visible in total signal intensities \cite{ayuso1, rego1}, and (ii) chirality polarization (or ``beam bending''), where enantio-sensitivity appears primarily as a spatial/angle redistribution of signal \cite{ayuso2, rego1}. Predicting which regime occurs for a given multicolour, structured, or multi-beam driving field remains non-trivial because the relevant interference conditions live in an extended Fourier space (frequency, transverse wavevector/emission angle, orbital angular momentum (OAM), and relative phases). 

An approach that has recently entered the field of nonlinear optics is the interpretation in terms of wave beatings \cite{trines1, trines2, trines3}. This method, which was originally developed in plasma physics, has been successfully applied to harmonic generation in plasma and solids, and provides a prediction of the generated harmonics and their properties \cite{trines1, trines2}. Its ramifications stretch well beyond plasma physics, into nonlinear optics and even the symmetry theory of crystal scattering \cite{trines3}. In summary, the driving field is decomposed into driving modes corresponding to circularly polarised fundamental photon modes (i.e. modes with spin $\sigma = \pm 1$), which are given signed frequencies $\omega/\sigma$ rather than positive-definite frequencies $\omega$. Then, the harmonics appear as beatings of the fundamental modes. This explains the generation of odd-order harmonics in typical HHG, see Figure \ref{figure1}(b). In addition, if spatial properties are present as field phase variations in the transversal plane, such as in non-collinear configurations or vortex beams, which carry orbital angular momentum (OAM) \cite{allen}, the reciprocal coordinate to the spatial coordinate where such variation occurs (i.e. $k_\perp/\sigma$ in the case of non-collinear configurations and $\ell/\sigma$ in the case of vortex beams) can be used as an additional dimension in the Fourier space, which becomes two-dimensional, see Figure \ref{figure1}(c). Thus, the beat-wave approach allows for the prediction of the linear momentum, spin angular momentum and orbital angular momentum of the harmonics. In addition, the anisotropy of the medium in a laser-solid interaction can be included in the beat-wave model as a zero-frequency (DC) driving mode \cite{trines1}

In this paper, we introduce a beat-wave description of chiral harmonic generation that (a) represents the enantio-sensitive response of the medium as a chiral DC mode, (b) classifies harmonic pathways by the parity (odd/even) of their DC‑mode contributions, and (c) reduces the global-chirality question to a simple closure condition on an ``odd'' sum of beat-step vectors in the relevant Fourier space. We also show three ways in which a specific laser beam configuration can be changed from global to local chirality and back. We validate this framework by applying it to several published configurations of synthetic chiral light, including bicircular OAM fields \cite{mayer1} and crossing multicolour beams \cite{ayuso1,ayuso2,rego1}.

\section{The chiral DC mode}

A (locally) chiral electromagnetic field is one whose electric‑field vector E(t) traces a trajectory in time that is not mirror-symmetric, i.e. changes handedness under an odd symmetry of the tangent space (mirror, improper rotation, 3-D inversion). A chiral molecule is a molecule whose mirror image is not identical to the molecule itself, and will thus also change handedness under an odd symmetry. This concept is closely related to that of a pseudoscalar (also known as an alternating 3-form or volume form) $f_3 (\vec{v}_1, \vec{v}_2, \vec{v}_3) \propto (\vec{v}_1 \times \vec{v}_2) \cdot \vec{v}_3$, which obeys the same symmetries.

To study harmonic generation in chiral molecules, we start from the pseudoscalar theory as given in e.g. Ayuso \emph{et al.} \cite{ayuso1}. For third-order sum frequency generation, where $\vec{E}_1 (\omega_1)$ and $\vec{E}_2 (\omega_2)$ combine to produce $\vec{E}_3 (\omega_1 + \omega_2)$ , one defines the third order field pseudoscalar $h^{(3)}$:
\[
h^{(3)}(-\omega_1 -\omega_2, \omega_1, \omega_2) \equiv (\vec{E}_1 \times \vec{E}_2) \cdot \vec{E}_3^*.
\]
Synthetic chiral light with nonzero $h^{(3)}$ requires three different non-co-planar frequency components. For synthetic chiral light with just two frequencies $\omega$ and $2\omega$, $h^{(3)} = 0$ and the lowest order nonzero pseudoscalar is $h^{(5)}$:
\[
h^{(5)} (-2\omega, -\omega, \omega, \omega, \omega) = \vec{E}_2^* \cdot (\vec{E}_1 \times \vec{E}_1^*) (\vec{E}_1 \cdot \vec{E}_1 ).
\]
In both pseudoscalars, we encounter the cross product of two non-collinear EM waves, so it makes sense to define this in terms of our beat-wave approach. Let $\vec{E}_1$ and $\vec{E}_2$ be two non-collinear EM waves with planar linear polarisation, and let $\Psi_i \equiv \exp [ i(\omega_i t - \vec{k}_i \cdot \vec{x} )/\sigma_i ]$ denote the ``fundamental modes'' that make up these waves. Then $\vec{E}_1 \times \vec{E}_2^*$ is s-polarised, and a corresponding function $\Psi_{1,2}$ can be defined as follows:
% Finally, if p-polarised pump beams with non-equal frequencies are used \cite{ayuso1,neufeld1,vogwell1}, one will have to resort to a more general expression for $\Psi_\mathrm{DC}$, such as:
\begin{equation}
\label{eq:crossproduct}
\Psi_{1,2} \equiv (\vec{E}_1 \times \vec{E}_2^*)_s = \sum_{i,j} \Psi_i \Psi_j^* (\sigma_j/\omega_j) c^2 (\vec{k}_i \times \vec{k}_j )_s/(\omega_i \omega_j).
\end{equation}
This function has the property that two Fourier modes with opposite $(\omega/\sigma, \vec{k}/\sigma)$ will have phases that are $\pi$ apart. This will be important when dealing with the case of two p-polarised laser beams with the same frequency $\omega$ crossing at a narrow angle \cite{ayuso1,ayuso2,rego1,neufeld1,vogwell1}. In this same case, $\Psi_{1,2}$ will be time-independent and return the perpendicular spin density $S_\perp$, as discussed in more detail below. For time-independent $\Psi_{1,2}$, we will use $\Psi_\mathrm{DC} \propto \Psi_{1,2}$, to make the connection with our earlier work \cite{trines1,trines2}.

The simplest configuration, as used in various publications \cite{ayuso1,ayuso2,rego1}, is composed of two fields at frequency $\omega$ and a third at frequency $2\omega$. Such a field can be created by using a focused field at $\omega$ with a polarisation in the direction of focus, e.g. both polarisation and focus in $\vec{e}_x$ (a line focus with the line along $\vec{e}_y$), or two crossing beams with $x$-polarisation and small transverse wave numbers $\pm k_x$ \cite{ayuso1, ayuso2, rego1}, or both polarisation and focus in $\vec{e}_r$ (point focus). In such configurations, the condition that $\nabla\cdot\vec{E} = 0$ in vacuum guarantees the existence of a longitudinal component $E_z$, and then $\vec{E}_1(\omega) \times \vec{E}_2(\omega) \propto E_x E_z \vec{e}_y$ or $\propto E_r E_z \vec{e}_\varphi$ respectively. This should then lead to even harmonics of $\omega$ in the $\vec{e}_y$ or $\vec{e}_\varphi$ direction. When overlaid with an external (achiral) field $\vec{E}(2\omega)$ in  $\vec{e}_y$ or  $\vec{e}_\varphi$ respectively, a non-trivial interference between the achiral $\vec{E}(2\omega)$ and the first even harmonic $E^2(\omega) E_x E_z \vec{e}_y$ or $E^2(\omega) E_r E_z \vec{e}_\varphi$ will result.

For two p-polarised driving modes at frequency $\omega$, as in Refs. \cite{ayuso1, ayuso2, rego1}, the chiral DC mode will have effective s-polarision. As in our earlier work \cite{trines2}, this means that, without the $2\omega$ light all odd harmonics of the $\omega$ light would be p-polarised like the pump wave, while all even harmonics would be s-polarised like the DC mode. In addition, all odd harmonics will have an even number of DC interactions and will thus be achiral, while all even harmonics will have an odd number of DC interactions and will thus be chiral. This is shown clearly in e.g. Figure 4 of Ayuso \emph{et al.} \cite{ayuso1} or Figure 7 of Rego and Ayuso \cite{rego1}. If this configuration is then overlaid with an achiral external $2\omega$ field with s-polarisation, the s-polarised light will then contain both chiral and achiral even harmonics, whose interference pattern will depend on the handedness of the chiral molecule under consideration. This way, the handedness of an enantio-pure sample or the ratio of L- and R-molecules in a mixture can be diagnosed. While this discussion is appropriate for a line focus, it can easily be applied to a point focus if one uses polar coordinates $(r,\varphi)$, polarisation and focusing in $\vec{e}_r$ and thus a DC mode with effective polarisation in $\vec{e}_\varphi$.

We note that a configuration with a ``line focus'' would be the ``continuum limit'' of configurations with two (or more) beams crossing at a narrow angle. We would like to emphasise this, to highlight the trajectory from two crossing beams to a line focus (both with a DC mode in $\vec{e}_y$) to a point focus (DC mode in $\vec{e}_\varphi$).

The concept of interference between chiral and achiral harmonics is illustrated by e.g. Rego \& Ayuso \cite{rego1}, where we find the following two expressions for chiral and achiral polarisations (somewhat reformatted):
\begin{align}
\vec{P}_c(2N\omega) &\propto [\vec{F}(\omega) \cdot \vec{F}(\omega)]^{N-1}  [\vec{F}(\omega) \cdot \vec{F}(\omega)] [\vec{F}^*(\omega) \times \vec{F}(\omega) ], \\
\vec{P}_a(2N\omega) &\propto [\vec{F}(\omega) \cdot \vec{F}(\omega)]^{N-1} \vec{F}(2\omega).
\end{align}
From this, it is clear that the main interference between chiral and achiral polarisations is between $\vec{F}(2\omega)$ and $[\vec{F}(\omega) \cdot \vec{F}(\omega)] [\vec{F}^*(\omega) \times \vec{F}(\omega) ]$; compare this to the equation for $h^{(5)}$ above.
% as an extension to the quantity $\vec{E}_3 \cdot (\vec{E}_1 \times \vec{E}_2)$ that we saw before.

For a plane wave with electric field $\vec{E} = \exp(i\omega t) (\vec{e}_x + i\varepsilon\vec{e}_y)/\sqrt{2}$, we find that $\vec{E}^* \times \vec{E} = i\varepsilon \vec{e}_z$ (spin pseudovector). This invites the identification of $\Im[\vec{F}^*(\omega) \times \vec{F}(\omega) ]$ with the spin density $\vec{S}_\omega$ of the $\omega$-field. However, for a purely plane wave, $\vec{S}$ is parallel to the direction of propagation and thus perpendicular to $\vec{E}$; a focused wave (or two p-polarised waves crossing at a narrow angle) is needed for interference between $\vec{S}$ and $\vec{E}$. We define $\vec{S}_\perp$ as the component of the spin density perpendicular to the wave propagation, which can interfere with $\vec{E}$. In addition, the factors $ [\vec{F}(\omega) \cdot \vec{F}(\omega)]$ invite identification with the ``beat steps'' in our earlier beat-wave approach to laser harmonic generation \cite{trines1, trines2, trines3}.
% Yes, this can be beat-waved.
% P_a: achiral harmonics using only pump laser modes, or maybe an even number of chiral DC contributions. Oh, Ayuso et al. missed that last one!
% P_c: chiral harmonics using an odd number of chiral DC contributions, could be larger than 1 !

For a focused field, $\vec{S}_\perp$ can be calculated as follows. We use a linearly polarised field $\vec{E}.= (E_x, 0, E_z)$ as an example, with $E_x = f(x) \cos(\omega t - k z)$. We use $\nabla \cdot \vec{E} = 0$ in vacuum to obtain $E_z = [f'(x)/k] \sin(\omega t - k z)$ and $S_y = [(f^2)'/(2k)]$. This means that (i) For a field focused in the direction of polarisation, $\vec{S}_\perp$ will be perpendicular to both propagation and polarisation directions, e.g. $E_x$ focused in $x$ will contribute to $S_y$; (ii) If the directions of focusing and polarisation are perpendicular, then there is no contribution to $\vec{S}_\perp$, e.g. $E_y$ focused in $x$ will not contribute to $\vec{S}_\perp$; (iii)  $S_y$ will have twice the effective transverse wave number of $\vec{E}$ and (iv) the zeroes of $S_y$ correspond to maxima or minima of $f^2(x)$, i.e. the spin density ``pattern'' is always $90^\circ$ out of phase with the intensity pattern of $\vec{E}$.

Along similar lines, one can calculate $S_\varphi$ for a field with a point focus. We note, however, that $S_\varphi$ will exhibit an effective $\ell/\sigma = -1$. while $S_y$ for a line focus will exhibit $\ell/\sigma = 0$.

% I may have missed it, but is C defined? e.g. C is a pseudoscalar coupling proportional to the enantio‑sensitive susceptibility? The factor (nL−nR) encodes the enantiomeric excess?

For a single-frequency field $\vec{E}_\omega$, the spin density $\vec{S}_\omega$ does not depend on time. We can thus use $\vec{S}_\omega$ to define a chiral DC mode, to use in a ``beat wave'' model for HHG in chiral media. For a focused driving field $\vec{E}$ with a transverse optical spin density $\vec{S}_\perp$, we therefore propose a DC mode along these lines:
\begin{equation}
\label{eq:dcmode}
\Psi_\mathrm{DC} \propto C(n_L - n_R) \vec{S}_\perp.
\end{equation}
Here, $C$ denotes a (chiral) molecular pseudoscalar: $C \not= 0$ for chiral molecules, while $C=0$ for achiral ones, and $C$ is even/odd under eve/odd isometric coordinate transformations; $n_{LR}$ denote the molecular densities for L- and R-molecules. This DC mode changes sign when the handedness of the chiral molecules is changed, and also when a single field component of the $\omega$-field is mirrored; thus, it has the right symmetry properties for the study of HHG in chiral media. If the curve described by the field vector $\vec{E}(\omega)$ remains in a single plane, then mirroring in this plane will not change $\Psi_\mathrm{DC}$; however, a field $\vec{E}(2\omega)$ perpendicular to that plane will change sign under such a mirror symmetry, and the quantity $\vec{E}(2\omega) \cdot \Psi_\mathrm{DC}$ will then change sign under any mirror symmetry in 3-D, as well as when the handedness of the molecule is changed. Quantities like $\vec{E}(2\omega) \cdot \Psi_\mathrm{DC}$ or $\vec{E}(2\omega) \cdot E^2(\omega) \Psi_\mathrm{DC} \equiv h^{(5)}$ play a central role in the study of chiral harmonic generation \cite{ayuso1, rego1}, and  can indeed be embedded in our beat-wave approach to HHG also.

If multiple pump beams are used, $\vec{S}_\perp$ and $\Psi_\mathrm{DC}$ may vary on short spatial scales due to interference between the pump beams. In that case, one should decompose $\Psi_\mathrm{DC}$ into modes whose amplitude $|\Psi_\mathrm{DC}|$ is constant on short spatial scales, i.e. the equivalent of pure circular polarisation. We will see an example of this below, when discussing chiral harmonic generation between two multicolour laser beams crossing at a small angle \cite{ayuso1, ayuso2, rego1}. 

The identification of the chiral DC mode and its properties via Eqns (\ref{eq:crossproduct}) and (\ref{eq:dcmode}) constitutes the first key result of this work. As in our previous work \cite{trines1, trines2, trines3}, understanding this DC mode is crucial to the understanding of the harmonic spectrum generated by laser beams interacting with a medium.

% Lerner et al. assume that $\Psi_\mathrm{DC} \to \Psi_\mathrm{DC}$ for any symmetry of the achiral pump field that maps $E^2 \to E^2$.
% This puts significant restrictions on the set of allowed DC modes. What are the implications of this?
% Lerner et al. got it right by accident for their simple case.

% Shape of the focal spot dictates which ``extra'' modes one gets and what the DC mode is.
% This dictates the set of symmetries of the focused beam, which may well be different from that of the original, unfocused, plane wave.
% Beware: set of symmetries of the field that generates the DC mode, e.g. $E_\omega$ only for the ``crossing beams'' case.
% Under this new set of symmetries, $E_\omega^2 \to E_\omega^2$ also implies that $\Psi_\mathrm{DC} \to \Psi_\mathrm{DC}$.
% Line focus: $\Psi_\mathrm{DC}$ preserved. Point focus: $\Psi_\mathrm{DC} \exp(-i\varphi)$ preserved???
% Makes sense, since Psi = E1 x E2 or whatever, so even if $E \to -E$ then still $\Psi_\mathrm{DC} \to \Psi_\mathrm{DC}$.
% Spin does not change sign under $E \to -E$.
% Extend to more generic E1.(E2 x E3). You are already looking at DC modes with nonzero wave vectors anyway. Also nonzero frequencies?

% Not obvious that Lerner et al. can distinguish between zero and nonzero amplitudes for $K=0$.
% Not obvious that they can distinguish between the two cases for the crossing-beams case: ``global chirality'' and ``beam bending''.
% Probably not, since these things are governed by the $2\onega$ light, which does not contribute to the DC mode for the ``crossing beams''.

\section{Conditions for global and local chirality}

Now that we have identified the chiral DC mode, generated in our case by the interaction of a focused beam at $\omega$ with a chiral molecule, we proceed to the study of chiral versus achiral harmonics, and their interference. In the beat-wave approach, a path to a harmonic takes the form $\Psi_i \prod_{j,k} (\Psi_j^* \Psi_k)$, where the functions $\Psi$ denote fundamental modes with pure circular polarisation. The chiral DC mode will change its sign when the handedness of the chiral molecule is changed, while the achiral external laser modes will of course not do this. Thus, a harmonic involving an odd number of factors $\Psi_\mathrm{DC}$ (usually one) will be chiral, while one involving an even number of factors $\Psi_\mathrm{DC}$ (usually zero) will be achiral.

Two harmonics with the same frequency $\omega/\sigma$ will have a static interference pattern that can be studied. If one is chiral and one is not, then the interference term $\Psi_a^* \Psi_c$ will be constant in time, but change sign when the handedness of the molecule is changed. This is exploited to determine the handedness of a chiral molecule, or to study the ratio of L- to R-molecules in a mixture. The function $h^{(5)}$ defined above is an example: interference between chiral and achiral second harmonic light. If an interference term $\Psi_a^* \Psi_c$ can be found that does not depend on any transverse coordinate ($x$, $y$, $r$, $\varphi$ or an emission angle), then the laser configuration is said to exhibit ``global chirality''; if all interference terms $\Psi_a^* \Psi_c$ depend on at least one transverse coordinate, then the laser configuration is said to exhibit only ``local chirality''.

As in our earlier work on symmetries, we define the vectors $X \equiv [t, x, y, z]$ and $K \equiv [\omega, -k_x, -k_y, -k_z]/\sigma$ and their inner product $K\cdot X \equiv (\omega t - \vec{k}\cdot\vec{x})/\sigma$. For two modes $\Psi_{A,B} = \exp( i K_{A,B}\cdot X)$, we find $\Psi_B^* \Psi_A = \exp[ i (K_A - K_B)\cdot X]$, and also $\Psi_\mathrm{DC}^* \Psi_A = \exp[ i (K_A - K_\mathrm{DC} ) \cdot X]$. We note that $(\Psi_\mathrm{DC}^* \Psi_B)^* (\Psi_\mathrm{DC}^* \Psi_A) = | \Psi_\mathrm{DC} |^2 \Psi_B^* \Psi_A$; if $\Psi_\mathrm{DC}$ is defined correctly so $| \Psi_\mathrm{DC} |^2$ is constant, we can define all cross terms $\Psi_B^* \Psi_A$ via cross-terms of the form $\Psi_\mathrm{DC}^* \Psi_A$, and thus in terms of difference vectors $K_A - K_\mathrm{DC}$.

We define the set $\{ K'_i \}$ of all the vectors $K'_i = K_a - K_\mathrm{DC}$, where $K_a$ is an achiral (pump) laser mode and $K_\mathrm{DC}$ is a chiral DC mode. We consider a chiral harmonic $\Psi_a$ and an achiral harmonic $\Psi_c$ with the same value of $\omega/\sigma$, so $\Psi_a^* \Psi_c$ is time-independent. (In practice, finite observation time and spectral resolution mean that interference is assessed within a finite frequency bin; we therefore restrict attention to chiral and achiral contributions that fall within the same resolved harmonic frequency.) Since $\Psi_c = \Psi_{c,i} \prod_{j,k} (\Psi_{c,j}^* \Psi_{c,k})$ and $\Psi_a = \Psi_{a,i} \prod_{j,k} (\Psi_{a,j}^* \Psi_{a,k})$, we find that $\Psi_a^* \Psi_c$ is purely a product of ``beat terms'' $\Psi_j^* \Psi_k$, and can be written as $\Psi_a^* \Psi_c = \exp[ i (\sum_j n_j K'_j) \cdot X]$. Because $\Psi_a^* \Psi_c$ must contain an odd number of factors $\Psi_\mathrm{DC}$ by construction, the sum $\sum_j n_j K'_j$ contains an odd number of terms, i.e. $\sum_j n_j $ is an odd integer. The $\omega/\sigma$ component of $\sum_j n_j K'_j$ will be zero by construction. 

% We then distinguish two cases. If there is a combination of $K'_j$ such that $\sum_j n_j K'_j = 0$ for $\sum_j n_j $ odd, then $\Psi_a^* \Psi_c$ is constant in both time and space and the chirality of the laser configuration is global. Conversely, if $\sum_j n_j K'_j $ is not zero for any $n_j$ with $\sum_j n_j $ odd, then $\Psi_a^* \Psi_c$ still has a spatial dependence and the chirality is only local.

We can now formulate the key result of our paper as follows. For an achiral laser mode $K_{a,i}$ and a chiral DC mode $K_\mathrm{DC}$ we define the chiral ``beat step'' vector $K'_i \equiv K_{a,i}-K_\mathrm{DC}$. Let $\{ K'_i \}$ be the set of all possible vectors $K'_i$. 
Then a chiral-achiral $\Psi_a^* \Psi_c$ interference term exists that is globally chiral (i.e. independent of all transverse coordinates) if there exist integers $n_i$ with $\sum_i n_i$ odd such that $\sum_j n_j K'_j = 0$. If no such odd closure exists in the chosen Fourier space, enantio-sensitivity appears only in coordinate‑dependent interference patterns (local chirality).

This leads us to our second key result: the question whether a given laser configuration is globally or locally chiral is reduced to the question whether a non-trivial ``odd'' sum of vectors from a given set can reach zero. 

\subsection{Phase as a coordinate}

At this point, one also needs to consider the role of phases of the fundamental modes in the laser-target configuration. Phases can determine both the direction of polarisation and a shift in time or position. For e.g.  $\Psi = \exp i(\omega t + \delta_+) + \exp i(-\omega t - \delta_-) = \cos[\omega t + (\delta_+ + \delta_-)/2 ] \exp i(\delta_+ - \delta_-)/2$, we get a polarisation rotation of $(\delta_+ - \delta_-)/2$ and a time advance of $(\delta_+ + \delta_-)/(2\omega)$. Pump beam phases feature prominently in various chiral laser configurations \cite{rego1}, so they should be included as coordinates. The question remains whether to include them as a component of $X$ or of $K$. We note that phases are a property of the beams, like $\omega$ or $\vec{k}$, rather than a property of spacetime, like $t$ or $\vec{x}$. Also, it is usually the phase differences that count rather than their absolute values; similar to the differences in $K$ that determine the beat steps, and unlike the specific values one uses for $X$. We therefore include controllable relative phases as components of $K$, since they are properties of the driving modes rather than spacetime coordinates. We implement this by adding a dummy coordinate $x_\delta \equiv 1$. With these definitions, we obtain $K \cdot X = (\omega t - \vec{k} \cdot \vec{x} - \ell\varphi - x_\delta \delta)/\sigma$ and once again $\Psi \propto \exp(i K\cdot X)$.

%Break the numbered list into 3 subheadings:
%
%1. Why treat phases as Fourier components.
% Phase is a beam property, not a spacetime property, and it's phase differences that count. Tight focus: in the EP configuration, the phase difference between the L- and R-modes stems from the angle of the major axis of the ellipse. Crossing beams: the phase difference stems from the relative phases of the 2w light, and how that relates to the phases of the DC modes, which are determined by the w light.
%2. Experimental meaning (CEP/relative phase control).
% Partly overlaps with the above?
%3. Effect of restricting a coordinate (slits / angular gating).
% One slit: global chirality. Two slits: beam bending. Range: may well be local chirality. You know. Now apply to the various examples.

For a phase difference $\Delta \delta$ and $0 \leq \Delta \delta \leq 2\pi$, the quantity $\cos(x_\delta \Delta\delta)$ will cover the full interval $[-1, 1]$. For smaller intervals for $\Delta\delta$, the cosine may cover a smaller interval also. Thus, the impact of chirality on the harmonic spectrum can be tuned via the range of $\Delta \delta$. 

For two crossing beams with global chirality: $\Delta \delta = 0$ and $\cos(x_\delta \Delta\delta) = 1$ in all directions, i.e. same harmonic spectrum in all directions. This value is chosen for maximum difference between L- and R-molecules. With local chirality: $\Delta \delta = 0$ or $\pi$ for the two directions and $\cos(x_\delta \Delta\delta) = \pm 1$ as a consequence. This results in different harmonic spectra for the two directions (beam bending); the directions change roles when L-molecules are swapped for R-molecules. For smaller differences in $\Delta \delta$ between the two directions, the difference between their spectra should be less dramatic but still visible.

For the crossing-beams case, the phase difference between chiral and achiral light is tied to the emission angle as well as to the phases of the $2\omega$ beams. This means that global chirality can be induced in a configuration with local chirality if the harmonic light is observed though a narrow slit, thus selecting  a narrow range of emission angles and thus for $\Delta \delta$ and $\cos(x_\delta \Delta\delta)$. In general, choosing a single value for $\cos(x_\delta \Delta\delta)$ corresponds to ``global chirality'' \cite{ayuso1}, two distinct values (e.g. for two distinct emission angles) to ``beam bending'' \cite{ayuso2} (provided that one angle corresponds to maximum emission for L-molecules while the other corresponds to maximum emission for R-molecules) and a full range of values to ``local chirality''. See also Section \ref{sec:switching}.

For the case of CP beams with  tight focus \cite{mayer1}: for $0 \leq \Delta \delta \leq 2\pi$, the full range of $\cos(x_\delta \Delta\delta)$ is indeed obtained. Specific values of $\Delta\delta$ can be picked to obtain the maximum difference between L- and R-molecules.

% ``Least bad'' model: define a ``phase coordinate'' $x_\delta \equiv 1$, then $\phi = \omega t - k x + \delta x_\delta$ or whatever.
% Note that for fixed $x$ and $0 \leq \Delta k \leq 2\pi/x$, the term $\pm \cos(x\Delta k)$ gives you everything you want from a chiral medium, yay.
% This basically forces to 	ALWAYS include any phase as a Fourier coordinate. Even crazier than I expected, but you do what you have to do.
% Tightly focused beams with EP pulse: three vectors with third component 0, 0 and $\delta$ respectively. Harmonics at same $(\omega/\sigma, \ell/\sigma)$ will be separated by $3\delta$ in the third coordinate.
% Two crossing beams: the phase of the DC modes is fixed at $\pm \pi/2$. If the phase of the $\2\omega$ light is also fixed at $\pm \pi/2$, then $\Delta \delta = 0$ and $\pm \cos(\Delta \delta)$ is fixed at $\pm 1$. You have redundancy and the phase term does what you want.
% If the phase of the $\2\omega$ light is fixed at $0$, then $\Delta \delta = \pm \pi/2$ and $\pm \cos(\Delta \delta) = 0$ while $\pm \sin(\Delta \delta) = \pm 1$. You have redundancy and either the sine or the cosine does what you want. You can make this work.
% For intermediate phases, there will be no ``official'' redundancy, but fluctuations in $\pm \cos(\Delta \delta)$ will be smaller. Also what you'd expect.
%
% Phase as a dimension in Fourier space: who would have thought it?

\section{Application to known configurations of synthetic chiral light}

\subsection{Tightly focused bicircular CP light with orbital angular momentum}

% Note that the DC mode here does not really have two components with phase difference $\pi$: the ``inner'' component is far stronger than the ``outer'' component, as opposed to the crossing-beams case, where the two components have equal strength.

We first study the configuration used by N. Mayer \emph{et al.} \cite{mayer1}, who use CP pulses with $\omega/\sigma = +1$ and $-2$ and OAM levels $\ell/\sigma = +1$
for both pulses. Expressions for the fields of a focused CP pulse with OAM can be found in e.g. Baumann and Pukhov \cite{baumann1}:
\begin{align}
E_r  &= -A_\perp \sin[\omega t - k z - (\ell+1)\varphi], \\
E_\varphi &= \sigma A_\perp \cos[\omega t - k z - (\ell+1)\varphi], \\
E_z &= [A_\perp/(k_0 r)] ( |\ell/\sigma| - \ell/\sigma -2r^2/w^2(z) )  \cos[\omega t - k z - (\ell+1)\varphi], \\
S_\varphi &= [A^2_\perp/(k_0 r)] ( |\ell/\sigma| - \ell/\sigma -2r^2/w^2(z) ).
\end{align}
Here, $A_\perp(r,z)$ is the appropriate envelope for a Laguerre-Gaussian beam with indices $\ell$ and $p=0$ and $w(z)$ is the beam waist. We note that both beams will contribute to the ``perpendicular spin'' $S_\varphi$. Since $\ell/\sigma = +1$ and thus $|\ell/\sigma| - \ell/\sigma = 0$ for both beams, their relative contributions to $S_\varphi$ will mostly be determined by their respective intensities and by how well the beam envelopes can be made to overlap in an actual experiment.

In the original configuration, we discern two achiral pump modes given by $K_A = (\omega_A/\sigma_A, \ell_A/\sigma_A) = (1,1)$ and $K_B = (-2, 1)$. The chiral DC mode is given by $K_\mathrm{DC} = (0,-1)$. This yields $K'_1 = (1,2)$ and $K'_2 = (-2,2)$. These two vectors are independent, so $n_1 K'_1 + n_2 K'_2 = 0$ implies $n_1 = n_2 = 0$ and thus $n_1 + n_2$ even. Global chirality is not possible in this case. We note that $2 K'_1 + K'_2 = (0,6)$, so the time-independent interference pattern for a given, constant value of $\omega/\sigma$ will have 6-fold azimuthal symmetry. This is shown by Mayer \emph{et al.} \cite{mayer1}, in their discussion of the function $h^{(5)} (r,\varphi)$ (which does indeed have 6-fold symmetry) and the interference patterns for various harmonic frequencies.

From $2 K'_1 + K'_2 = (0,6)$, we find that the ``interference term'' $h^{ (5) } (r,\varphi) \propto (\Psi_A \Psi^*_\mathrm{DC} )^2 \Psi_B \Psi^*_\mathrm{DC}$, so $h^{(5)}$ will change sign when switching from L- to R-molecules. This can be seen in Figure 2a-d by Mayer \emph{et al.} \cite{mayer1}: when switching from L- to R-molecules, the minima and maxima trade places. Integrating the interference pattern over $0^\circ \leq \varphi \leq 360^\circ$ will return the same value for both L- and R-molecules, since the fluctuations average out. 

In general, we find that $K'_1 = (\omega_A/\sigma_A, 1+ \ell_A/\sigma_A)$ and $K'_2 = (\omega_B/\sigma_B, 1+ \ell_B/\sigma_B)$, so $(\omega_B/\sigma_B) K'_1 - (\omega_A/\sigma_A) K'_2 = [0, (\omega_B/\sigma_B)(1+ \ell_A/\sigma_A) - (\omega_A/\sigma_A)(1+ \ell_B/\sigma_B) ]$. In the case that $n_1 (\omega_A/\sigma_A) + n_2 (\omega_B/\sigma_B) = 0$ with $n_1 + n_2$ odd, examples can be found where $n_1 K'_1 + n_2 K'_2 = (0,0)$. Thus, the configuration can be made globally chiral (function $h^{(5)} (r,\varphi)$ independent of $\varphi$)  for the right choice of parameters, as mentioned (but not discussed) by Mayer \emph{et al.} \cite{mayer1}. For example, if $K_A = (1,0)$ and $K_B = (-2, -3)$, we find that $2 K'_1 + K'_2 = 2(1,1) + (-2, -2) = (0,0)$. Similarly, for $K_A = (1,-1)$ and $K_B = (-2, -1)$, we find that $2 K'_1 + K'_2 = 2(1,0) + (-2, 0) = (0,0)$ also. This shows that (i) a change in parameters can change a configuration from locally to globally chiral and back, and (ii) how this is fully incorporated in our ``beat wave'' description of harmonic generation in chiral media.

Global chirality can be induced in the original configuration [$K_A = (1,1)$ and $K_B = (-2,1)$] if the $\ell/\sigma$ dimension can be removed from Fourier space. Fixing the azimuthal angle $\varphi$ at a specific value corresponds to integrating over all $\ell/\sigma$, which would remove that dimension. In this new, reduced situation, $K'_1 = (1)$,  $K'_2 = (-2)$ and $2 K'_1 + K'_2 = (0)$, so an odd number of $K'$ vectors adds up to zero, indicating global chirality. Again returning to Mayer \emph{et al.} \cite{mayer1}: if the far field in their results is observed through a narrow slit around $\varphi = 10^\circ$, R-molecules will yield maximum intensity through the slit while L-molecules will yield minimum intensity, indicating ``global'' chirality within the slit. A second slit positioned around $\varphi = 100^\circ$ will yield maximum (minimum) intensity for L-molecules (R-molecules). Thus, harmonic light from L-molecules will be ``bent'' towards $\varphi = 100^\circ$ while light from R-molecules will be ``bent'' towards $\varphi = 10^\circ$. 

We conclude: (i) When all angles $0^\circ \leq \varphi \leq 360^\circ$ are considered, this configuration is locally chiral, with a criterion for $K'_i$ to match; (ii) if only one specific value for $\varphi$ is chosen, the coordinate $\ell/\sigma$ is effectively eliminated from Fourier space and the configuration becomes effectively globally chiral, with a criterion for the reduced $K'_i$ to match; (iii) for two specific well-chosen values for $\varphi$, even beam bending (similar to Ayuso \emph{et al.} \cite{ayuso2}) can be induced.

Next, we consider the situation when the beam at $\omega$ is given elliptic polarisation instead of circular. This implies the addition of a third fundamental mode $K_C = (-1,-1)$, having a small amplitude, and a third vector $K'_3 = (-1,0)$. We note that $K'_1 - K'_2 + 3K'_3 = (0,0)$, so this new configuration is globally chiral: the function $h^{(5)}$ now contains a $\varphi$-independent contribution that will not vanish after integration over $\varphi$. Adding a third fundamental mode provided enough redundancy to achieve this. We also note that $K'_1 + K'_3 = K'_2 - 2K'_3 = (0,2)$, so $h^{(5)}$ will also contain a strong $\exp(2i \varphi)$ contribution and possibly a weaker $\exp(4i \varphi)$ , as signalled by Mayer \emph{et al.} \cite{mayer1}.

If we vary the phase $\delta$, we need to include it as a Fourier coordinate also. We obtain $K_C = (-1,-1,\delta)$ and also $K_\mathrm{DC} = (0,-1,0)$, $K_A = (1,1,0)$, $K_B = (-1, 1, 0)$, $K'_1 = (1,2,0)$, $K'_2 = (-2,2,0)$ and $K'_3 = (-1, 0, \delta)$. Since the $K'_i$ are now independent, this extended configuration will not show full global chirality; the response between L- and R-molecules should change with the phase $\delta$, as shown in Figure 4 of Mayer \emph{et al.} \cite{mayer1}. The harmonic spectrum is given by $K = K_\mathrm{DC} + \sum_i n_i K'_i$. For $\ell/\sigma = 0$ and $\omega/\sigma \equiv -1\ (\mathrm{mod\ } 3)$, the leading-order dependence on $\delta$ should vary as $3\delta$, otherwise as $\delta$. The paper by Mayer \emph{et al.} \cite{mayer1} does not contain the necessary data to study this (harmonics grouped by $\omega$ rather than $\omega/\sigma$), but it would be an interesting topic to investigate.

\subsection{Two-colour laser beams focusing at a narrow angle}

This is a complex, many-layered case. This is reflected in the number of papers published on it: global chirality \cite{ayuso1}, beam bending \cite{ayuso2}, a summary paper \cite{rego1}, SFG vs THG \cite{vogwell1}, and so on. There is even an apparent conflict between Ayuso \emph{et al.} \cite{ayuso1}  and Neufeld, Tzur and Cohen \cite{neufeld1} on one hand, and Lerner \emph{et al.} \cite{lerner1} on the other. We will treat it in stages.

First stage: two p-polarised LP beams at $\omega$ crossing at a narrow angle. Propagation in $z$, in-plane field is mainly in $x$, out-of-plane field (if any) in $y$. Then $E_x \propto \cos(k_x x) \cos(\omega t - k_z z)$, $E_z \propto \sin(k_x x) \sin(\omega t - k_z z)$, spin $S_y \propto \sin(2 k_x x)$. We set the phases of the $\omega$ beams to be zero, so those of the DC modes are $\pm \pi/2$. Vectors in $(\omega/\sigma, k_\perp/\sigma, \delta/\sigma)$ space: $\pm (1, \pm k_x, 0)$ for the $\omega$  beams and $\pm (0, 2k_x, \pi/2)$ for the DC modes. Since the $\omega$ beams are p-polarised and the DC modes are effectively s-polarised, this configuration will (to leading order) generate achiral odd harmonics with p-polarisation and chiral even harmonics with s-polarisation. On its own, there will be no chiral-achiral interference because the achiral and chiral harmonics have orthogonal polarisations and different frequencies. However, if an achiral external $2\omega$ beam with s-polarisation is added, then achiral even harmonics with s-polarisation can be generated, as well as chiral odd harmonics with p-polarisation. Achiral-chiral interference will now be ubiquitous, and can be controlled via the phase differences between the various pump beams.

Second stage: Vogwell \emph{et al.} \cite{vogwell1} introduce a single $2\omega$ beam with s-polarisation and $K = \pm(2, 2k_x, \phi_{2\omega} )$, with $\phi_{2\omega}$ to be tuned for maximum effect. They also use a DC mode which is not just time-independent but also space-independent, i.e. $k_\perp/\sigma = 0$ rather than $k_\perp = \pm 2k_x$. Thus, $K_\mathrm{DC} = (0,0,0)$ (different from Ayuso \emph{et al.} \cite{ayuso1}, see also below), $K_{AB} = \pm (1, \pm k_x, 0)$ and $K_C = \pm (2, 2k_x, \phi_{2\omega})$, so $K'_{1,2} = (1, \pm k_x, 0)$ and $K'_3 = (2,  2k_x, \phi_{2\omega} )$. Without either $K'_1$ or $K'_3$, the remaining two vectors are independent and the configuration will show only local chirality. With all three vectors present and $\phi_{2\omega}=0$ fixed, we find e.g. $K'_3 - 2K_1 = 0$ (odd number of vectors), so global chirality is possible.

Achiral $3\omega$ light: Paths to $(3,3)$, $(3,1)$, $(3,-1)$ and $(3,-3)$ involving only $\omega$ light are all third order. Modes $(3,1)$ and $(3,-1)$ will have the highest intensity; modes $(3,3)$ and $(3, -3)$ are possible but will be less intense. Achiral paths that also involve $2\omega$ light are fifth order and will not be considered further. Chiral $3\omega$ light: $(3,1) = (1,-1) + K'_3$, which corresponds to $\Psi_\omega \Psi_{2\omega} \Psi^*_\mathrm{DC}$ and is thus third order; reaching $(3,-1)$ requires at least two contributions of $K'_2$, so there are no third order chiral paths to this harmonic. Lowest order chiral path: $(3,-1) = (1,-1) - K'_1 + K'_2 + K'_3$, which corresponds to $\Psi^*_A \Psi^2_B\Psi_C \Psi^*_\mathrm{DC}$ and is thus fifth order. In short the mode $(3,1)$ can be reached via a chiral and an achiral path that are both third order and have a decent amplitude, so this mode is the best option. This is clearly borne out by Vogwell \emph{et al.} \cite{vogwell1}.

If the phase $\phi_{2\omega}$ is kept fixed at an optimal value, global chirality is obtained. If two values are used, e,g, $\phi_{2\omega} = 2\pi/3$ and $\phi_{2\omega} = 5\pi/3$, then $3\omega$ light from an L-molecule is ``bent'' towards $\phi_{2\omega} = 2\pi/3$ while light from an L-molecule is ``bent'' towards $\phi_{2\omega} = 5\pi/3$. If the harmonic light is considered for all values of $\phi_{2\omega}$ then only ``local chirality'' is found as the intensity fluctuates with $\phi_{2\omega}$ for both L- and R-molecules.

Third stage: Ayuso \emph{et al.} \cite{ayuso1} use two sets of two-colour crossing beams: two beams at $\omega$ with p-polarisation, overlaid with two beams at $2\omega$ and s-polarisation. The dependence of the field on both $t$ and $x_\perp$ is considered. For the $\omega$ field, we assume that $x=0$ and the phase $\delta = 0$ for a maximum in the transverse envelope. The field description is as follows:
\begin{align}
E_x &\propto \cos(k_x x) \cos(\omega t - k_z z),\\
E_z &\propto \sin(k_x x) \sin(\omega t - k_z z),\\
E_y &\propto \cos[2 k_x x + (\phi_+ -\phi_-)/2 ] \cos[2\omega t - 2k_z z + (\phi_+ +\phi_-)/2 ],\\
S_y &\propto \sin(2 k_x x).
\end{align}
From this, also using the phase $\delta/\sigma$ as a dimension in Fourier space, we obtain the following fundamental modes in $(\omega/\sigma, k_\perp/\sigma, \delta/\sigma)$ space: $K_\mathrm{DC} = \pm (0, 2, \pi/2)$ (i.e. dependent on $x$), $K_A = \pm (1,1,0)$, $K_B = \pm (1,-1,0)$, $K_C = \pm (2, 2, \phi_+)$ and $K_D = \pm (2, -2, \phi_-)$. We note the difference with Vogwell \emph{et al.} \cite{vogwell1}, who use $K_\mathrm{DC} = (0,0,0)$ (not dependent on $x$). Similar to Ayuso \emph{et al.} \cite{ayuso1}, we first set $\phi_+ = -\phi_- = \pi/2$, so $K_C = \pm (2, 2, \pi/2)$ and $K_D = \pm (2, -2, -\pi/2)$. Among others, we find $K'_1 = (1,-1, -\pi/2)$, $K'_2 = (1,1, +\pi/2)$ and $K'_3 = (2,0,0)$. Since $K'_1 + K'_2 - K'_3 = (0,0,0)$ (odd number of vectors), there will be global chirality, in agreement with Ayuso \emph{et al.} \cite{ayuso1} or Rego \& Ayuso \cite{rego1}.

Fourth stage: following Ayuso \emph{et al.} \cite{ayuso2} we set $\phi_+ = \phi_- = 0$, so $K_C = \pm (2, 2, 0)$ and $K_D = \pm (2, -2, 0)$; this yields $K'_3 = (2,0,-\pi/2)$ and $K'_4 = (2,0,\pi/2)$, among others. In any event, it follows that the third component of any $K'$ vector equals $\pm \pi/2$, so no odd sum of $K'$ vectors will ever return $(0,0,0)$; an even sum is always needed. Thus, this configuration will always show local chirality: fluctuations ``even out'' when integrated over all emission angles. However, if one concentrates on a single emission angle, the emission will be maximum for e.g. L-molecules and minimum for R-molecules, while the situation will be reversed for the opposite emission angle (``beam bending'' \cite{ayuso2}). These findings are in full agreement with Ayuso \emph{et al.} \cite{ayuso2} or Rego \& Ayuso \cite{rego1}.

If we compare the third and fourth stages, we see that the configuration can be changed from ``globally chiral'' to ```locally chiral'' via a change in the parameters of the pump beams, in this case the relative phases of the s-polarised $2\omega$ beams. The consequence of this change in relative phase is an increase or decrease in the number of dimensions of the span of the set $\{ K'_i \}$. Fewer dimensions imply global chirality, while more dimensions imply local chirality.

% [If a single $\omega$ beam and a single $2\omega$ beam are crossed at a narrow angle, this will also yield interesting results depending on whether the dependence on the transverse coordinate $x$ is included in the calculations or not; see below.]

% Global chirality
% Local chirality
% THG with an ``incomplete'' setup, e.g. Vogwell et al.

\section{Switching chirality on and off}
\label{sec:switching}

Chirality: study the interference pattern between a chiral and an achiral harmonic. The chiral harmonic will change sign when switching between L- and R-molecules, the achiral harmonic will not. Let $\Delta K$ be the difference between a chiral and a nearby achiral harmonic (at least with the same $\omega/\sigma$, if possible also the same $\vec{k}/\sigma$), and let $X$ be the coordinate dual of $\Delta K$, with $0 \leq |X| \leq L$. The interference pattern will look like this: $I = a \pm b\cos(X\cdot \Delta K)$ for $0 \leq |X| \leq L$, $a, b, L > 0$. We write $\langle I \rangle \equiv (1/L) \int I(X) dX$. We distinguish three cases:
\begin{enumerate}
\item Global chirality: $\Delta K = 0$ or $L|\Delta K| \ll 1$, so $\langle I \rangle \approx a \pm b$.
\item Local chirality: $L|\Delta K| \gg 1$, so $\langle I \rangle \approx a$.
% \item Intermediate: $L|\Delta K| = \mathcal{O}(1)$, so $\langle I \rangle = a \pm [b/(L|\Delta K|)] \sin(L|\Delta K|)$.
\item Intermediate: $L|\Delta K| = \mathcal{O}(1)$, so $\langle I \rangle = a \pm b\sinc(L|\Delta K|)$.
\end{enumerate}
So far, we have concentrated on situations where $\Delta K = 0$, but we will now also consider scenarios where $L$ is small, which can equally induce global chirality.

Consequences for local chirality in general. For a wave number difference $\Delta (k/\sigma)$, it is often assumed that the transverse coordinate ranges over a length $L$ such that $L\Delta (k/\sigma) \leq 2\pi$, so $\cos[L\Delta (k/\sigma)]$ attains its full range. However, for smaller $L$, the cosine covers only part of this range; in particular, for $L\Delta (k/\sigma) \ll 1$, $\cos[L\Delta (k/\sigma)]$ is nearly constant. We note that a full range of $\cos[L\Delta (k/\sigma)]$ is associated with ``local chirality'', while a limited range is associated with ``global chirality''. While global chirality has so far been associated with $\Delta (k/\sigma) = 0$, the same result can be obtained via $L \ll 2\pi/ \Delta (k/\sigma)$.

Since the transverse laser pulse envelope in the far field is the Fourier transform of the envelope in the near field, one can use the same reasoning for Fourier coordinates like the emission angle or a phase difference: a single value of such a coordinate corresponds to ``global chirality'', two distinct values to ``beam bending'' and a full range of values to ``local chirality''. This can be exploited to induce global chirality by observing the harmonic light through a narrow slit, effectively eliminating a transverse coordinate.

Methods to induce global chirality concentrate on either reducing the dimension of the available Fourier space or introducing new modes and thus more vectors $K'_I$:
\begin{enumerate}
\item Adding a pump mode, e.g. by simply adding a laser beam or changing a beam's polarisation from circular to elliptic or linear \cite{mayer1}.
\item Fixing a spatial coordinate (or reducing it to a narrow range). This can be viewed as either setting $L\Delta K \ll 1$ or reducing the dimension of the Fourier space via eliminating the dual of the fixed spatial coordinate.
\item Changing the parameters of the beams, causing the same number of vectors to occupy fewer dimensions in Fourier space. In the ``tight focus' case \cite{mayer1}, this happens when $(\omega_B/\sigma_B)(1 + \ell_A/\sigma_A) - (\omega_A/\sigma_A)(1 + \ell_B/\sigma_B) = 0$. In the ``crossing beams'' case \cite{ayuso1, ayuso2, rego1}, this happens when the phase difference between the $2\omega$ beams is changed: matching the phases of the $2\omega$ light to those of the DC modes ensures that $\delta/\sigma$ is proportional to $k_x/\sigma$ for all $K'_i$ vectors, effectively eliminating the $\delta/\sigma$ dimension.
\item The work by Rego \& Ayuso \cite{rego1} can actually illustrate all three cases of chirality: global, local and intermediate. The amplitude of the interference pattern between chiral and achiral modes is $\propto 2|\cos[(\phi_+ - \phi_-)/2]|$ or similar. From this, one can establish a `` degree of global chirality'', given by $2|\sin[(\phi_+ - \phi_-)/2]|$ or similar. (i) $\phi_+ - \phi_- = \pi$: No fluctuations, fully globally chiral. (ii) $\phi_+ - \phi_- = 0$: it's all fluctuations, only local chirality. (iii) $0 < \phi_+ - \phi_- < \pi$: a mixture of the two. See also Sections 2.3 and 2.4 of Rego \& Ayuso \cite{rego1}.
% This is all related to the fact that the transverse range is ``sub-period'': if the transverse range of emission angles were unlimited, then there would only be the cases $\phi_+ - \phi_- = \pi$ (globally chiral) and $\phi_+ - \phi_- < \pi$ (locally chiral). But the ``phase coordinate'' is fixed: $x_\delta = 1$, so the intermediate case is also possible. Worth pointing that out at some level; ask Rego \& Ayuso for input.
\end{enumerate}
Methods to go back from global to local chirality are of course the opposite from the above.

How to restrict a (transverse) coordinate to one specific value, in order to induce global chirality: criteria for choosing an appropriate coordinate, as inspired by our work on ``directional frequency combs'' \cite{trines2}.
\begin{enumerate}
\item Start from a configuration that is not yet globally chiral. That means that no odd number of ``chiral'' steps will ever add up to zero. 
\item Thus, paths with odd and even numbers of steps can never end up at the same harmonic, or you'd be able to join them to make an odd path to zero (which we just ruled out).
\item Take a path involving an odd number of chiral steps. If you collapse the spectrum in that direction, you'll obtain global chirality.
\item Now take a path which involves an even number of chiral steps, and which is not an (even) integer multiple of an``odd'' path. Collapsing in that direction will not get you global chirality.
\item The directions under 3 and 4 are always distinct, so those provide all the necessary criteria.
\item Collapse: effectively projecting onto the space $K_c \cdot X = 0$, e.g. via fixing the coordinate dual of $K_c$.
\end{enumerate}

For example: in the case of the tightly focused OAM beams by Mayer \emph{et al.} \cite{mayer1}, we find that $K_c = 2 K'_1 + K'_2 = (0, 6)$ which is an odd (i.e chiral) path connecting a chiral and an achiral mode with the same $\omega/\sigma$. Projection onto the space $K_c \cdot X = 0$ corresponds to eliminating $\varphi$ (the coordinate dual of $\ell$) and retaining only $t$ as an independent coordinate. Thus, fixing $\varphi$ to a specific value will turn this configuration from locally chiral into globally chiral, as borne out by the results of Mayer \emph{et al.} \cite{mayer1}.

Thus, our third key result is the identification of three ways to induce global chirality in a locally chiral laser configuration: (i) adding a fundamental laser mode, to increase redundancy; (ii) changing the laser parameters to decrease the number of dimensions of the span of the set of vectors $\{ K'_i \}$; (iii) restricting a spatial coordinate to a narrow range, to decrease the number of dimensions of the configuration's Fourier space.

\subsection{Chiral dichroism allowed or not: an example in practice}

In this section, we discuss a case where various groups claim to study the same laser beam configuration but reach opposite conclusions regarding the chirality of this setup. See Ayuso \emph{et al.} \cite{ayuso1}, supplemental material; Neufeld, Tzur and Cohen \cite{neufeld1}; Lerner \emph{et al.} \cite{lerner1}, supplemental material; Vogwell \emph{et al.} \cite{vogwell1}.

Configuration: a chiral molecule irradiated by elliptically polarised light at $(\omega, -k_x)$ and $(2\omega, 2k_x)$. Ayuso \emph{et al.} \cite{ayuso1} or Neufeld, Tzur and Cohen \cite{neufeld1} state that their simulations indicate global chirality (chiral dichroism). Lerner \emph{et al.} \cite{lerner1} state that chiral dichroism is forbidden for this configuration, according to their  symmetry theory,

This apparent contradiction can be resolved by observing that Ayuso \emph{et al.} do not consider any dependence on the transverse coordinate $x$, while Lerner \emph{et al.} do. So Lerner \emph{et al.} use $K'_1 = (1,-1)$ and $K'_2 = (2,2)$, which are independent, so no odd sum of these will ever yield $(0,0)$. Thus, Lerner \emph{et al.} will not find global chirality for their specific situation. However, Ayuso \emph{et al.} use only one Fourier dimension $(\omega/\sigma)$: $K'_1 = (1)$ and $K'_2 = (2)$ and $2K'_1 - K'_2 = (0)$ (odd number). So Ayuso \emph{et al.} will find global chirality for their specific situation. It may appear that these two groups are studying the same problem, but deep down they are not. Also, the fact that ``collapsing'' the $x$-coordinate will change a configuration from ``local CD'' to ``global CD'' supports our findings.

Studying the symmetries of $E^2$ for the electric field $\vec{E}$ as used by the two groups is also revealing.
Neufeld, Tzur and Cohen use $E_x = \cos(\omega t)  + \cos(2\omega t)$ or similar, which does not depend on $x$. In this case, $E^2$ has a single discrete even symmetry: $t \to t + 2\pi/\omega$, while $E \to E$ under this symmetry. Conversely, Lerner \emph{et al.} use $E_x = \cos(\omega t + k_x x)  + \cos(2\omega t - 2k_x x)$ or similar. They study this symmetry of $E^2$: $(t,x) \to [t + 3\pi/(4\omega),\ x - \pi/(4 k_x)]$, while $E \to -E$ under this symmetry. 
Conclusion: these two groups are dealing with rather different collections of symmetries, and since the symmetries define the problem, they are effectively solving two rather different problems. So it could be expected that they would reach different conclusions.

\section{Conclusions}

In this paper, we have developed a beat-wave approach to laser harmonic generation in chiral media. This approach is based on three key results. The first key result is the derivation of a chiral DC mode to complement the achiral laser pump modes to generate the full harmonic spectrum of both chiral and achiral harmonics. Interference between a chiral and an achiral harmonic can then be used to diagnose the chirality of the medium.

As our second key result, we have derived a beat‑wave criterion for global chirality: global enantio‑sensitive interference is possible if an odd integer combination of chiral beat‑step vectors $K'_i$ (which encode the difference between a chiral DC mode and an achiral pump laser mode) closes to zero in the relevant extended Fourier space: $(\omega/\sigma, \vec{k}/\sigma, \ell/\sigma, \delta/\sigma)$.

If only even integer combinations of chiral beat-step vectors will close to zero, then the configuration will be locally chiral, and the interference pattern between chiral and achiral modes will depend on at least one transverse coordinate.

As our third key result, we have derived three ways in which a locally chiral configuration can be made globally chiral, which are all aimed at increasing the redundancy of the set $\{K'_i\}$.  (i) Add a fundamental mode to the pump laser configuration, to increase the number of vectors $K'_i$; (ii) adjust the parameters of the pump laser modes, so the span of the set $\{K'_i\}$ loses a dimension; (iii) if the chiral-achiral interference pattern depends on a specific transverse coordinate,  fix this oordinate to a specific value, so the vectors $K'_i$ lose a non-trivial dimension and the span of the set $\{K'_i\}$ loses a dimension.

In more detail:
\begin{enumerate}
\item ``Chirality'' identified in terms of interference between chiral and achiral paths to the same harmonic mode. Chiral paths include an odd number of contributions from a chiral DC mode $\Psi_\mathrm{DC}$. Achiral paths include an even number of such contributions, or none at all.
\item Identification of the chiral DC mode $\Psi_\mathrm{DC} \propto \exp(i _\mathrm{DC} \cdot X)$ , usually in terms of the spin density $\vec{S} \propto \vec{E}(\omega) \times \vec{E}^*(\omega)$. This requires a focused field with a nonzero ``transverse'' spin density $\vec{S}_\perp$. For a line focus, one finds e.g. $\vec{S}_\perp = S_\perp \vec{e}_y$ \cite{ayuso1, ayuso2, rego1}, while for a point focus one finds $\vec{S}_\perp = S_\perp \vec{e}_\varphi$ \cite{mayer1}.
% For a 3-D chiral field, the $\vec{E}_\omega$ field must have a longitudinal component $E_z$ in addition to its transverse components $E_{x,y}$, and its spin density must have a transverse component $S_\perp$ in addition to its longitudinal component $E_z$. This requires focusing of the $\vec{E}_\omega$ field. The DC mode is then proportional to the transverse spin density $S_\perp$.
\item Identification of ``odd'' steps $K'_i = K_X - K_\mathrm{DC}$, where $K_X$ is some achiral pump mode. Identification of ``odd'' paths to a harmonic (odd number of odd steps) and ``even'' paths (even number of such steps, or none at all).
\item If both an odd and an even path lead to the same harmonic, then (i) they can interfere to provide information about the handedness of the DC mode, and (ii) this means that $\Psi_c \Psi^*_a = 1$, i.e there is an odd number of odd steps $K'_i$ that adds up to zero. This provides a simple, general criterion to determine whether or not a given configuration of laser modes will exhibit chiral dichroism in a chiral medium.
\item We have demonstrated how the phase differences between pump modes can be introduced as coordinates in extended Fourier space, which is necessary for the study of certain complex laser beam configurations.
\end{enumerate}

We have applied our new criteria to (i) tightly focused Laguerre-Gaussian beams with circular polarisation \cite{mayer1}, $\omega$-$2\omega$ beams crossing at a narrow angle with $x$-dependence \cite{ayuso1, ayuso2, rego1} and without $x$-dependence \cite{vogwell1}. In each case, we can explain their findings qualitatively using our new model, demonstrating the versatility of our beat-wave approach to laser harmonic generation.

Our findings demonstrate how versatile our beat-wave approach to harmonic generation is. The generic criteria for global vs local chirality that we have developed, and the steps to change a configuration from one to the other, will advance the analysis of existing configurations of synthetic chiral light and inform the design of future configurations in both theory and experiment.

\section*{Acknowledgements}

This work was supported by EPSRC (grants EP/Z535692/1and EP/V049232/1). DA acknowledges funding from the Royal Society
URF\textbackslash{R}\textbackslash{251036}.
LR acknowledges that the project leading to these results has received funding from ``la Caixa'' Foundation (ID 100010434), under the agreement ``LCF/BQ/PR24/12050018'', the Spanish Ministry of Science, Innovation and Universities and the State Research Agency through the project ref. PID2024-163024NA-I00 (MICIU/AEI/10.13039/501100011033/FEDER, UE) and from the Severo Ochoa Centers of Excellence program through Grant CEX2024-001445-S.

\end{document}